\documentclass[aps,prd,preprint,groupedaddress,nofootinbib,showpacs]{revtex4-1}
\usepackage{mathrsfs,amsbsy,amssymb,latexsym,amsfonts,amsmath,amsthm}
\usepackage{graphicx,color}
\makeatletter
\catcode`\@=11
\def\@seccntformat#1{\csname the#1\endcsname.~~}
\makeatother
\newcommand{\nn}{\nonumber}

\begin{document}


\title{Matter fields in triangle-hinge models}


\author{Masafumi Fukuma}%
 \email{fukuma@gauge.scphys.kyoto-u.ac.jp}
\author{Sotaro Sugishita}%
 \email{sotaro@gauge.scphys.kyoto-u.ac.jp}
\author{Naoya Umeda}%
 \email{n\_umeda@gauge.scphys.kyoto-u.ac.jp}
\affiliation{Department of Physics, Kyoto University, Kyoto 606-8502, Japan}

\date{\today}

\begin{abstract}

The worldvolume theory of membrane 
is mathematically equivalent to three-dimensional quantum gravity 
coupled to matter fields corresponding to the target space coordinates 
of embedded membrane. 
In a recent paper \cite{Fukuma:2015xja}
a new class of models are introduced 
that generate three-dimensional random volumes, 
where the Boltzmann weight of each configuration 
is given by the product of values 
assigned to the triangles and the hinges.  
These {\em triangle-hinge models} describe three-dimensional pure gravity 
and are characterized
by semisimple associative algebras. 
In this paper, we introduce matter degrees of freedom to the models 
by coloring simplices in a way that they have local interactions. 
This is achieved simply by extending the associative algebras 
of the original triangle-hinge models, 
and the profile of matter field is specified by the set of colors 
and the form of interactions.  
The dynamics of a membrane in $D$-dimensional spacetime 
can then be described by taking the set of colors to be $\mathbb{R}^D$.  
By taking another set of colors, 
we can also realize three-dimensional quantum gravity 
coupled to the Ising model, the $q$-state Potts models or the RSOS models. 
One can actually assign colors to simplices of any dimensions 
(tetrahedra, triangles, edges and vertices), 
and three-dimensional colored tensor models can be realized 
as triangle-hinge models 
by coloring tetrahedra, triangles and edges at a time.

\end{abstract}


\maketitle


\section{Introduction}
\label{intro}
M-theory is a candidate for the theory of everything including quantum gravity, 
where membranes are believed to be fundamental objects 
\cite{Hull:1994ys,Townsend:1995kk,Witten:1995ex,Schwarz:1995jq,Horava:1995qa}. 
The worldvolume theory of membrane is mathematically equivalent 
to three-dimensional quantum gravity coupled to matter fields 
corresponding to the target space coordinates of embedded membrane 
\cite{Polyakov:1987ez}.  
However, our understanding of three-dimensional quantum gravity coupled to matter 
is still not at a sufficient level  
if we compare it with the two-dimensional case, 
where  the dynamics of random surfaces has been well understood 
from various perspectives 
by using matrix models as an analytic tool 
(see, e.g., \cite{DiFrancesco:1993nw} for a review). 
In fact, matrix models generate random surfaces as Feynman diagrams 
and can be solved analytically. 
This solvability enables us to find a critical point 
around which the continuum limit is taken, 
and we now have a clear understanding  
of two-dimensional quantum gravity coupled to a large class of matter fields 
(e.g., $c \leq 1$ noncritical string theories).  
Thus, we expect that our understanding of the dynamics of membranes 
will be substantially developed 
if we can find a three-dimensional analog of matrix models, 
which generates three-dimensional random volumes as Feynman diagrams 
and allows us to investigate the dynamics analytically 
(hopefully at the level of matrix models). 
It will then lead to a consistent formulation of M-theory 
if such models admit the introduction of supersymmetry 
and do not have an issue like the $c=1$ barrier in two-dimensional theories.   

Recently, as a first step towards this direction, 
the authors constructed a new class of models 
that generate three-dimensional random volumes \cite{Fukuma:2015xja}.  
Since the dynamical variables are given by matrices 
and each model can be specified by a semisimple associative algebra,  
these models have a potential to be solved analytically using matrix model techniques. 
We call these models  {\em triangle-hinge models} 
because each Feynman diagram is treated as consisting of  
``triangles and hinges.''%
\footnote{
A similar approach was taken for three-dimensional topological lattice field theories 
\cite{Chung:1993xr}. 
}
This is in sharp contrast to the setup 
in tensor models \cite{Ambjorn:1990ge, Sasakura:1990fs, Gross:1991hx} 
or in group field theory \cite{Boulatov:1992vp, Freidel:2005qe}, 
where the minimum unit of Feynman diagram is given by a tetrahedron. 
Triangle-hinge models have an intrinsic problem 
that three-dimensional volumes cannot be assigned 
to a large portion of Feynman diagrams. 
However, one can reduce the set of possible diagrams 
such that they represent only and all of the tetrahedral decompositions 
of three-dimensional manifolds, 
by introducing specific interaction terms 
and taking an appropriate limit of parameters in the models \cite{Fukuma:2015xja}. 
Therefore, triangle-hinge models can be regarded 
as discrete models of three-dimensional quantum gravity.  

The original triangle-hinge models given in \cite{Fukuma:2015xja}  
do not have any extra degrees of freedom 
other than those of simplicial decompositions  
and thus describe three-dimensional {\em pure} gravity. 
However, in order to describe the dynamics of membrane, 
we need to extend the models 
so that they contain matter fields corresponding to the target space coordinates. 

The main aim of this paper is to introduce local matter degrees of freedom 
to triangle-hinge models, 
by coloring simplices in tetrahedral decompositions 
[actually simplices of arbitrary dimensions 
(tetrahedra, triangles, edges and vertices)]. 
The coloring is realized 
within the algebraic framework of the original triangle-hinge models, 
and we only need to extend the defining semisimple associative algebras 
and to modify the interaction terms accordingly. 
The matter fields thus obtained have local interactions 
because colored simplices interact only with their neighbors.  

A matter field is specified 
by the set of colors and the form of interactions. 
The worldvolume theory of membrane 
is given by taking the set of colors to be $\mathbb{R}^D$ 
with a local interaction in the target spacetime.
Besides this, 
we can construct various spin systems on random volumes.  
For example, 
the Ising model on random volumes can be realized 
by taking the set of colors to be $\mathbb{Z}_2=\{+,-\}$ 
and by assigning a color ($\pm$) to each tetrahedron. 
We can also set up the $q$-state Potts models, the RSOS models \cite{Itzykson:1989sx}
and even more generic models on random volumes.  
We will further show that three-dimensional colored tensor models 
\cite{Gurau:2009tw}%
\footnote{
Although the original tensor models can generate diagrams 
not homeomorphic to pseudomanifolds, 
colored tensor models are free from this issue \cite{Gurau:2010nd}. 
Furthermore, it is known that colored tensor models 
have good analytical properties (see, e.g., \cite{Gurau:2011xp} for a review).
} 
can be realized as triangle-hinge models 
by assigning specific matter degrees of freedom 
to tetrahedra, triangles and edges at a time. 

This paper is organized as follows. 
In section \ref{review}, we review the basic structure of triangle-hinge models. 
In section \ref{coloring}, we give a general prescription 
to introduce matter degrees of freedom to the models.  
In section \ref{ctm}, 
we review the Feynman rules of colored tensor models 
and show that they can be reproduced from triangle-hinge models 
by coloring tetrahedra, triangles and edges in a specific way.  
Section \ref{conclusion} is devoted to conclusion.

\section{Review of triangle-hinge models}
\label{review}

\begin{figure}[htbp]
\begin{center}
\includegraphics[height = 3.3cm]{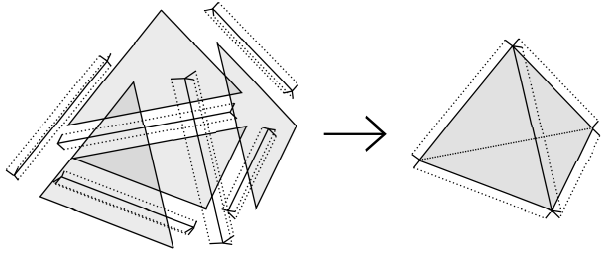}
\begin{quote}
\caption{
Construction of tetrahedral decompositions 
with triangles and multiple hinges \cite{Fukuma:2015xja}. 
}
\label{fig:tetra_triangle-hinge}
\end{quote}
\vspace{-6ex}
\end{center}
\end{figure}

In this section, we give a brief review of triangle-hinge models, 
which generate random diagrams 
consisting of triangles glued together along multiple hinges 
(see the original paper \cite{Fukuma:2015xja} for details). 
Note that a tetrahedral decomposition can always be regarded 
as a Feynman diagram of a triangle-hinge model 
as can be understood from Fig.~\ref{fig:tetra_triangle-hinge}. 

\subsection{Generalities}

We first give the definition of triangle-hinge models. 
The dynamical variables are given by $N\times N$ real symmetric matrices $A$ and $B$: 
\begin{align}
&A_{ij} = A_{ji}, \quad B^{ij} = B^{ji} \quad (i, j=1,\ldots, N), 
\label{sym_AandB} 
\end{align}
and the action takes the form 
\begin{align}
 S[A, B] = & \frac{1}{2} A_{ij}B^{ij}
 - \frac{\lambda}{6}\, C^{ijklmn}A_{ij}A_{kl}A_{mn} \nn \\
 & - \sum_{k \geq 2} \frac{\mu_k}{2k}\, B^{i_1 j_1} \cdots B^{i_k j_k}
 y_{i_1 \ldots i_k} y_{j_k \ldots j_1},  
\label{original_action}
\end{align}
where $C^{ijklmn}$, $y_{i_1\ldots i_k}$, 
$\lambda$ and $\mu_k$ are real-valued coupling constants. 
The Feynman diagrams are obtained 
by expanding the action \eqref{original_action} 
around the ``kinetic term'' $(1/2) A_{ij}B^{ij}$. 
The interaction vertices corresponding to 
$\lambda\, C^{i_1j_1i_2j_2i_3j_3}$ 
and  $\mu_k\, y_{i_1 \ldots i_k}\, y_{j_k \ldots j_1}$ 
can be represented by 
triangles and $k$-hinges, respectively, as in Fig.~\ref{fig:triangle-hinge}, 
if we assume the coupling constants to have the following symmetry properties: 
\begin{figure}[htbp]
\begin{center}
\includegraphics[height = 2.7cm]{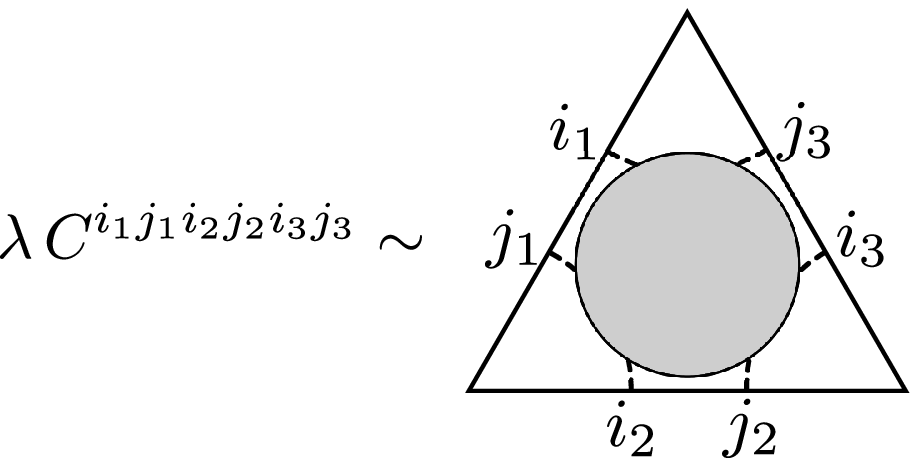}
\hspace{0.5cm}
\includegraphics[height = 2.7cm]{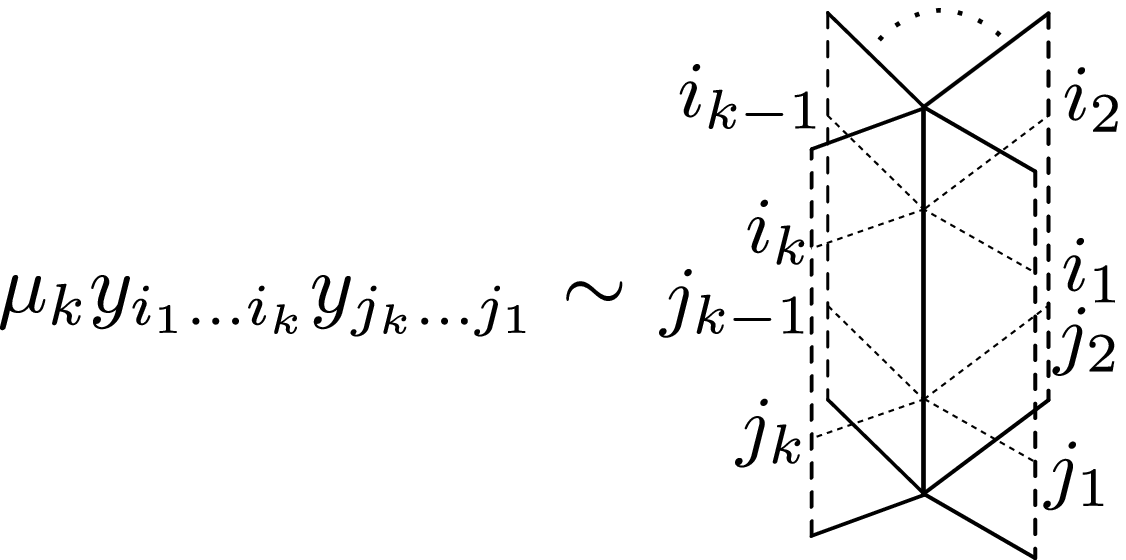}
\caption{
Triangles and multiple hinges \cite{Fukuma:2015xja}. 
}
\label{fig:triangle-hinge}
\vspace{5ex}
\end{center}
\end{figure}
\begin{align}
 &C^{i_1j_1i_2j_2i_3j_3}=C^{i_2j_2i_3j_3i_1j_1},  \quad
 C^{i_1j_1i_2j_2i_3j_3}=C^{j_3i_3j_2i_2j_1i_1},
\label{C_cycle}
\\
 &y_{i_1 i_2 \ldots i_k}=y_{i_2 \ldots i_k i_1}, 
\label{y_cycle}
\end{align}
which realize the symmetries of triangles and hinges 
under rotations and flips.%
\footnote{
In fact, when multiplied by $A_{i_1j_1}A_{i_2j_2}A_{i_3j_3}$ $(A_{ij}=A_{ji})$, 
only such part of $C^{i_1j_1i_2j_2i_3j_3}$ survive 
that are invariant under interchanges of indices 
$i_\alpha$ and $j_\alpha$ $(\alpha=1,\ldots,3)$ 
and under permutations of three pairs of indices 
$(i_1 j_1)$, $(i_2 j_2)$ and $(i_3 j_3)$. 
Thus, one could assume the symmetry 
$C^{i_1 j_1 i_2 j_2 i_3 j_3}=C^{j_1 i_1 i_2 j_2 i_3 j_3}
=C^{i_2 j_2 i_1 j_1 i_3 j_3}$ 
in the action \eqref{original_action}. 
We, however, do not assume this symmetry 
and regard the contributions from $C^{i_1j_1i_2j_2i_3j_3}$, 
$C^{j_1 i_1 i_2 j_2 i_3 j_3}$ and $C^{i_2j_2i_1j_1i_3j_3}$ 
as giving different Feynman diagrams. 
This prescription enables us to interpret the interaction vertices as triangles  
and is commonly adopted in the standard treatment of matrix models. 
Note that only the fully symmetric part is actually left 
when all the diagrams are summed.  
The same argument is applied to the hinge parts.
}  
The propagator has the form 
\begin{align}
 \langle A_{ij}B^{kl} \rangle = \delta_i^{\phantom{i}k} \delta_j^{\phantom{j}l}
 + \delta_i^{\phantom{i}l} \delta_j^{\phantom{j}k} 
 \label{prop}, 
\end{align}
where the two terms on the right-hand side correspond to two ways 
of gluing an edge of a triangle to that of a hinge (in the same or opposite direction). 
Thus,  the action \eqref{original_action} gives Feynman diagrams 
consisting of triangles 
which are glued together along multiple hinges in all possible ways. 

\begin{figure}[htbp]
\begin{center}
\includegraphics[height = 2.7cm]{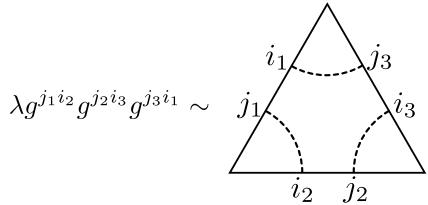}
\caption{
Index lines on a triangle \cite{Fukuma:2015xja}. 
}
\label{fig:triangle_indices}
\vspace{-3ex}
\end{center}
\end{figure}

A wide class of triangle-hinge models can be defined by semisimple associative algebras 
$\mathcal{A}$ of linear dimension $N$ \cite{Fukuma:2015xja}. 
With a basis $\{e_i\}$ $(i=1,\ldots,N)$ of $\mathcal{A}$ 
$\bigl[\mathcal{A}=\bigoplus_{i=1}^N \mathbb{R}\, e_i\bigr]$, 
the multiplication is expressed as 
\begin{align}
 e_i \times e_j = y_{ij}^{\phantom{ij}k}e_k.
\end{align}
Then, the cyclically symmetric rank $k$ tensor $y_{i_1 \ldots i_k}$ are constructed 
from the structure constants $y_{ij}^{\phantom{ij}k}$ as
\begin{align}
 y_{i_1 \ldots i_k} &\equiv y_{i_1 j_1}^{\phantom{i_1 i_1}j_k} 
 y_{i_2 j_2}^{\phantom{i_2 j_2}j_1} \ldots y_{i_k j_k}^{\phantom{i_k j_k}j_{k-1}}.
\label{subscript_y}
\end{align}
\newpage
\noindent
The rank two tensor $y_{ij}$ is especially denoted by $g_{ij}$ and is called metric, 
$g_{i j}\equiv y_{ij}=y_{i k}^{\phantom{j k}\ell}\,
y_{j \ell}^{\phantom{i \ell}k}$.%
\footnote{
An associative algebra $\mathcal{A}$ is semisimple 
(i.e.\ a direct sum of matrix rings) 
if and only if the metric $g=(g_{ij})$ has its inverse $g^{-1}\equiv (g^{ij})$ 
\cite{Fukuma:1993hy}. 
} 
A possible choice of $C^{ijklmn}$ satisfying \eqref{C_cycle} is
\begin{align}
 C^{ijklmn} =g^{jk}g^{lm}g^{ni},
\label{choice_C_g}
\end{align}
which corresponds to the index lines illustrated in Fig.~\ref{fig:triangle_indices}. 
This is not the unique solution to the condition \eqref{C_cycle}, 
and we will use this arbitrariness later [see \eqref{intro_omega}]. 

The free energy of the model is given by the summation 
of Boltzmann weights $w(\gamma)$ 
over all possible connected diagrams $\gamma$: 
\begin{align}
\log Z&=\sum_{\gamma} w(\gamma), \\
 w(\gamma) &= \frac{1}{S(\gamma)} \, \lambda^{s_2(\gamma)} 
 \Bigl( \prod_{k \geq 2} \mu_k^{s_1^k(\gamma)} \Bigr) \, \mathcal{F}(\gamma) , 
\end{align} 
where $S(\gamma)$ denotes the symmetry factor of diagram $\gamma$, 
$s_2(\gamma)$ the number of triangles, 
and $s_1^k(\gamma)$ the number of $k$-hinges. 
$\mathcal{F}(\gamma)$ is a function of $C^{ijklmn}$ and $y_{i_1 \ldots i_k}$ 
(and thus a function only of the structure constants $y_{ij}^{\phantom{ij}k}$) 
and is called the {\em index function of diagram} $\gamma$. 

\begin{figure}[htbp]
\begin{center}
\includegraphics[height = 3.0cm]{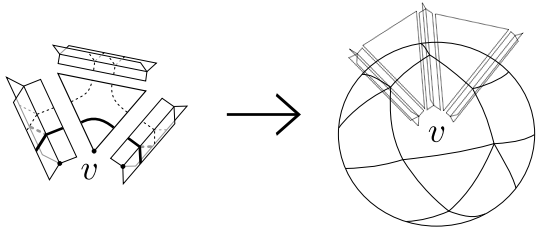}
\caption{
A part of index network (left) 
and a connected index network around vertex $v$ (right) \cite{Fukuma:2015xja}. 
When hinges share a common vertex $v$, 
their index lines are connected via intermediate triangles 
and give a polygonal decomposition of a closed surface enclosing vertex $v$. 
The closed surface needs not be a sphere, 
and we denote its genus by $g(v)$. 
}
\label{fig:index_network}
\vspace{-3ex}
\end{center}
\end{figure}

\begin{figure}[htbp]
\begin{center}
\includegraphics[height = 3.2cm]{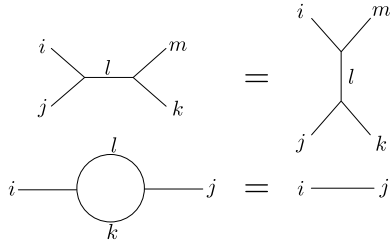}
\caption{
Fusion move (top) and bubble move (bottom) \cite{Fukuma:1993hy}. 
The index function $\zeta(v)$ is invariant 
under these two-dimensional topology-preserving local moves.  
}
\label{fig:associativity}
\vspace{-3ex}
\end{center}
\end{figure}

It is easy to see that 
the index function $\mathcal{F}(\gamma)$ is the product  
of the contributions $\zeta(v)$ from vertices $v$ 
(to be called the {\em index functions of vertices})\,: 
\begin{align}
 \mathcal{F}(\gamma) = \prod_{v:\,\text{vertex of $\gamma$}} \zeta(v) .
\end{align}
In fact, index lines out of different hinges are connected 
if and only if the hinges share a common vertex in $\gamma$, 
and then a connected component of index lines forms a polygonal decomposition of 
a closed two-dimensional surface enclosing a vertex 
(see Fig.~\ref{fig:index_network}).%
\footnote{
As is argued in \cite{Fukuma:2015xja}, 
a two-dimensional surface can be uniquely assigned 
to each connected index network by carefully following the contraction of indices.
} 
Moreover, $\zeta(v)$ is a two-dimensional topological invariant 
of the closed surface around $v$. 
In fact, $\zeta(v)$ is the product of $y_{ij}^{\phantom{ij}k}$ 
whose indices are all contracted appropriately, 
and is invariant under two-dimensional topology-preserving local moves 
that are generated by the fusion move and the bubble move 
(see Fig.~\ref{fig:associativity}), 
which are equivalent to the condition of associativity 
$y_{ij}^{\phantom{ij}l} y_{lk}^{\phantom{lk}m}
=y_{jk}^{\phantom{jk}l} y_{il}^{\phantom{il}m}$ 
and the definition of metric, $ y_{i k}^{\phantom{j k}l}\,
y_{j l}^{\phantom{i l}k}=g_{ij}$, respectively 
 \cite{Fukuma:1993hy}. 
Thus the index function $\zeta(v)$ is 
the two-dimensional topological invariant associated with algebra $\mathcal{A}$ 
\cite{Fukuma:1993hy}
and is characterized only by the genus $g(v)$ 
of the closed surface around vertex $v$,
$\zeta(v)=\mathcal{I}_{g(v)}[\mathcal{A}]$. 
Therefore, the free energy of the model takes the form
\begin{align}
 \log Z&=\sum_{\gamma} \frac{1}{S(\gamma)} \, \lambda^{s_2(\gamma)} 
 \Bigl( \prod_{k \geq 2} \mu_k^{s_1^k(\gamma)} \Bigr) \, 
\prod_{v:\,\text{vertex}} \mathcal{I}_{g(v)}[\mathcal{A}] 
\label{FreeEnergy}. 
\end{align}

\subsection{Matrix ring}

\begin{figure}[htbp]
\begin{center}
\includegraphics[height = 3.0cm]{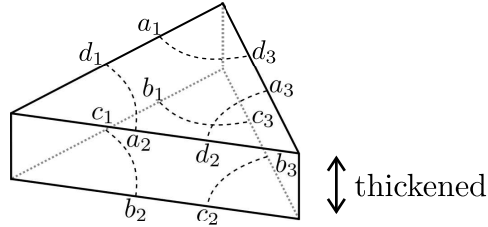}
\caption{
Index lines on a triangle in the case of matrix ring \cite{Fukuma:2015xja}. 
}
\label{fig:tri_fig}
\vspace{-3ex}
\end{center}
\end{figure}

\begin{figure}[htbp]
\begin{center}
\includegraphics[height = 3.0cm]{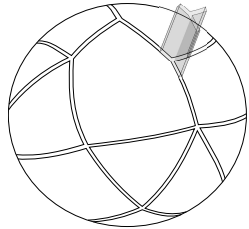}
\caption{
A connected index network with double lines \cite{Fukuma:2015xja}. 
This represents a polygonal decomposition 
of a closed surface. 
Each polygon will be called an index polygon. 
}
\label{fig:index_surface_double_line}
\vspace{-3ex}
\end{center}
\end{figure}

The simplest example of semisimple algebra is matrix ring 
$M_n(\mathbb{R})=\bigoplus\mathbb{R} e_{ab}$ 
(with linear dimension $N=n^2$). 
Here, we take the basis to be $\{e_{ab}\}$ $(a,b =1,\ldots n)$, 
where $e_{ab}$ is a matrix unit whose $(c,d)$ element is 
$(e_{ab})_{cd} =\delta_{ac} \delta_{bd}$. 
Note that indices $i$ are now double indices, $i=(a,b)$. 
When we take $\mathcal{A}=M_n(\mathbb{R})$ 
as the defining associative algebra of a triangle-hinge model, 
the choice of \eqref{subscript_y} and \eqref{choice_C_g} 
gives the action of the form \cite{Fukuma:2015xja}
\begin{align}
 S & = \frac{1}{2} A_{abcd} B^{abcd}
 - \frac{\lambda}{6n^3} A_{bacd} A_{dcef} A_{feab} \nn \\
&~~~ - \sum_{k \geq 2} \frac{n^2 \mu_k}{2k}
 B^{a_1 a_2 b_2 b_1} B^{a_2 a_3 b_3 b_2}
 \cdots B^{a_k a_1 b_1 b_k} .
\label{action_mat_ring}
\end{align}
Here, the variables $A$ and $B$ satisfy 
\begin{align}
 A_{abcd}=A_{cdab}, \quad B^{abcd}=B^{cdab},
\label{mat_sym}
\end{align}
and we have used the fact that 
the tensor 
$C^{i_1j_1i_2j_2i_3j_3}=C^{a_1 b_1 c_1 d_1 a_2 b_2 c_2 d_2 a_3 b_3 c_3 d_3}$ 
in \eqref{choice_C_g} 
takes the form
\begin{align}
 C^{a_1 b_1 c_1 d_1 a_2 b_2 c_2 d_2 a_3 b_3 c_3 d_3} 
 = \frac{1}{n^3}\, \delta^{d_1 a_2} \delta^{d_2 a_3} \delta^{d_3 a_1}
 \delta^{b_3 c_2} \delta^{b_2 c_1} \delta^{b_1 c_3}.
\label{C_matrix}
\end{align}
The interaction terms can then be expressed by thickened triangles 
as in Fig.~\ref{fig:tri_fig}. 
Accordingly, index lines in Fig.~\ref{fig:index_network} 
are written with double (or thickened) lines as in Fig.~\ref{fig:index_surface_double_line}. 
Polygons formed by index loops 
will be called {\em index polygons}. 
One can show that 
$\mathcal{I}_g$ in \eqref{FreeEnergy} is given by $n^{2-2g}$ 
for a connected index network of genus $g$ \cite{Fukuma:2015xja}. 

Furthermore, the model with $\mathcal{A}=M_n(\mathbb{R})$ 
has a duality 
which interchanges the roles of triangles and hinges \cite{Fukuma:2015xja}.
In fact, with the new variables dual to $A$ and $B$:%
\footnote{
We will use this duality transformation 
when we discuss a duality of coloring in subsection \ref{0simplex_coloring}. 
} 
\begin{align}
\tilde{A}_{abcd} \equiv A_{bcda}, \quad \tilde{B}^{abcd} \equiv B^{bcda},
\label{duality_trsf}
\end{align}
the action \eqref{action_mat_ring} can be rewritten to the form
\begin{align}
 S = & \frac{1}{2} \tilde A_{abcd} \tilde B^{abcd}
 - \frac{\lambda}{6n^3} \tilde A_{abcd} \tilde A_{befc} \tilde A_{eadf} \nn \\
& - \sum_{k \geq 2} \frac{n^2 \mu_k}{2k} \tilde B^{a_1 b_1 b_2 a_2}
 \tilde B^{a_2 b_2 b_3 a_3} \cdots \tilde{B}^{a_k b_k b_1 a_1} .
\label{action_mat_dual}
\end{align}
The way to contract the indices of $\tilde{A}$ (or $\tilde{B}$) 
in the dual action \eqref{action_mat_dual} 
is the same as that of $B$ (or $A$) 
in the original action \eqref{action_mat_ring}. 
Thus, in the dual picture, the diagrams consist of polygons and 3-hinges, 
which are actually the dual diagrams to the original ones.

\subsection{Restriction to tetrahedral decompositions} 
\label{reduction_manifolds}

The diagrams generated in the model \eqref{action_mat_ring} 
consist of triangles whose edges are randomly glued together, 
and generally do not represent tetrahedral decompositions. 
However, one can define models 
such that the leading contributions in a large $N=n^2$ limit 
represent (only and all of the) tetrahedral decompositions. 
By denoting the defining associative algebra by $\mathcal{A}_{\rm grav}$, 
this can be achieved 
by (i) taking $\mathcal{A}_{\rm grav}$ to be 
$M_{n=3m}(\mathbb{R})$ 
with $n$ a multiple of three, 
(ii) modifying the tensor 
$C^{a_1 b_1 c_1 d_1 a_2 b_2 c_2 d_2 a_3 b_3 c_3 d_3}$ 
from \eqref{C_matrix} 
to%
\footnote{
This modification can be absorbed into a modification of the kinetic term 
by redefining $A_{abcd}$ as
$\omega^{d'a}A_{abcd} \, \omega^{bc'} \to A_{d'c'cd}$. 
One then can show that there still exists a duality between triangles and hinges. 
} 
\begin{align}
 C_{\rm grav}^{a_1 b_1 c_1 d_1 a_2 b_2 c_2 d_2 a_3 b_3 c_3 d_3} 
 \equiv \frac{1}{n^3}\, \omega^{d_1 a_2} \omega^{d_2 a_3}
 \omega^{d_3 a_1} \omega^{b_3 c_2} \omega^{b_2 c_1} \omega^{b_1 c_3}
\label{intro_omega}
\end{align}
with a permutation matrix $\omega$ of the form
\begin{align}
\omega \equiv 
\begin{pmatrix} 0&1_{n/3}&0 \\ 0&0&1_{n/3} \\ 1_{n/3}&0&0 \end{pmatrix},
& \qquad 1_m : m\times m \, \text{unit matrix},
\end{align}
and (iii) taking an appropriate limit of parameters in the model 
\cite{Fukuma:2015xja}. 

\begin{figure}[htbp]
\begin{center}
\includegraphics[height = 3.5cm]{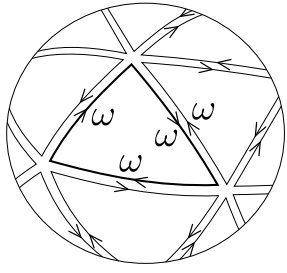}
\caption{
A connected index network with double lines 
in the presence of matrix $\omega$ \cite{Fukuma:2015xja}. 
}
\label{fig:index_surface_double_line_omega}
\vspace{-3ex}
\end{center}
\end{figure}

In fact, with this modification, 
each index polygon with $\ell$ segments 
gives a factor $\mathrm{tr}\,\omega^\ell$, 
which vanishes unless $\ell\equiv 0$ $(\mbox{mod}~3)$ 
(see Fig.~\ref{fig:index_surface_double_line_omega}). 
Thus, the index function $\zeta(v)=\mathcal{I}_{g(v)}$ at vertex $v$ 
takes a nonvanishing value ($=n^{2-2g(v)}$) 
only when the number of segments of every index polygon 
is a multiple of three. 
As proved in \cite{Fukuma:2015xja} in detail, 
the possible number of segments can be further reduced to three 
by taking the limit $n\to\infty$ 
with $n^2\, \mu_k$ and $n/\lambda$ being fixed,  
and there are left only such diagrams that represent tetrahedral decompositions.%
\footnote{
The set of possible diagrams can be further reduced 
such as to represent three-dimensional {\em manifolds} 
by introducing a parameter to control the number of vertices \cite{Fukuma:2015xja}. 
} 

\section{Introducing matter degrees of freedom}
\label{coloring}

The above prescription to reduce the configurations 
to tetrahedral decompositions also works 
when $\mathcal{A_{\rm grav}}$ is extended 
to a tensor product of the form 
$\mathcal{A} = \mathcal{A}_{\rm grav} \otimes \mathcal{A}_{\rm mat}$. 
Here, $\mathcal{A}_{\rm grav}$ is again $M_{n=3m}(\mathbb{R})$, 
and $\mathcal{A}_{\rm mat}$ is another semisimple associative algebra 
to be characterizing matter degrees of freedom. 
In fact, since the structure constants of $\mathcal{A}$ 
are given by the product of the structure constants 
of $\mathcal{A}_{\rm grav}$ 
and those of $\mathcal{A}_{\rm mat}$, 
the index function $\mathcal{F}(\gamma)$ of each diagram $\gamma$ 
is factorized to the product of the contributions 
from $\mathcal{A}_{\rm grav}$ and $\mathcal{A}_{\rm mat}$
if we set the tensor $C$ to take a factorized form $C=C_\mathrm{grav} C_\mathrm{mat}$: 
\begin{align}
 \mathcal{F}(\gamma) \equiv \mathcal{F}(\gamma;\mathcal{A})
 = \mathcal{F}(\gamma;\mathcal{A}_{\rm grav})\, 
 \mathcal{F}(\gamma; \mathcal{A}_{\rm mat})
 \equiv \mathcal{F}_{\rm grav}(\gamma)\,\mathcal{F}_{\rm mat}(\gamma).
\label{F_factored}
\end{align}
Then, by setting $C_{\rm grav}$ to the form \eqref{intro_omega} 
and by taking the limit $n\to\infty$ 
with $n^2\, \mu_k$ and $n/\lambda$ being fixed as in \ref{reduction_manifolds}, 
the index function $\mathcal{F}(\gamma)$ vanishes 
unless $\gamma$ represents a tetrahedral decomposition, 
and thus we can reduce the set of possible diagrams 
to tetrahedral decompositions 
independently of the choice of $\mathcal{A}_{\rm mat}$.%
\footnote{
Note that the introduction of matter degrees of freedom 
may further reduce the set of possible diagrams 
because $\mathcal{F}_{\rm mat}(\gamma)$ may vanish for a subset of simplicial decompositions. 
} 
In this section, 
assuming that this reduction is already made, 
we show that a set of colors (representing matter degrees of freedom) 
can be assigned to simplices of arbitrary dimensions 
(tetrahedra, triangles, edges and vertices) 
by choosing $\mathcal{A}_{\rm mat}$ and interaction terms appropriately. 

Note that 
for $\mathcal{A}=\mathcal{A}_{\rm grav}\otimes\mathcal{A}_{\rm mat}$ 
the dynamical variables take the form $A_{abcd,ij}$ and $B^{abcd,ij}$, 
where $(ab,cd)$ are matrix indices for $\mathcal{A}_{\rm grav}$ 
and $(i,j)$ are those for $\mathcal{A}_{\rm mat}$. 
In the rest of paper, 
we omit the indices $a,b,\ldots$ 
with respect to $\mathcal{A}_{\rm grav}$ 
in order to simplify expressions.  
We will denote the set of colors by $\mathcal{J}$ 
and the number of elements by $|\mathcal{J}|$.

\subsection{Coloring tetrahedra}
\label{3simplex_coloring}

\begin{figure}[htbp]
\begin{center}
\includegraphics[height = 3.0cm]{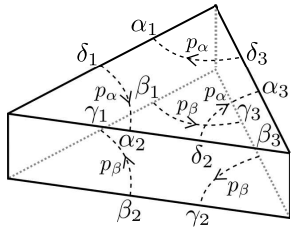}
\caption{
A thickened triangle. 
The upper (lower) side has color $\alpha$ ($\beta$). 
}
\label{fig:3simplex_coloring_fig}
\end{center}
\end{figure}

\begin{figure}[htbp]
\begin{center}
\includegraphics[height = 3.7cm]{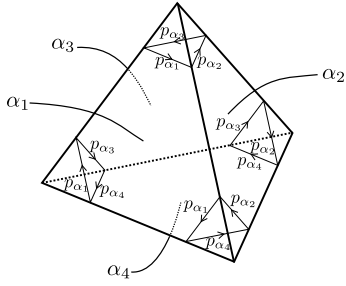}
\caption{
Index triangles inside a tetrahedron with triangles colored 
as in \eqref{3simplex_coloring_int}. 
}
\label{fig:tetra_index_triangle_fig}
\vspace{-6ex}
\end{center}
\end{figure}

We show that tetrahedra can be colored 
despite the fact that the action \eqref{original_action} does not have 
interaction terms corresponding to tetrahedra. 
We first set 
$\mathcal{A}_{\rm mat}=M_{|\mathcal{J}|}(\mathbb{R})
=\bigoplus_{\alpha,\beta \in \mathcal{J}}\mathbb{R}\,e_{\alpha\beta}$ 
and let the interaction terms take the form%
\footnote{
Recall that we are only looking at the matter part. 
Actually, the variable $A$ has extra indices of 
$\mathcal{A}_{\rm grav}=M_{n=3m}(\mathbb{R})$ 
as $A_{abcd,\alpha\beta\gamma\delta}$, 
and the interaction terms \eqref{3simplex_coloring_int} 
have extra factors \eqref{intro_omega}.
\label{fn_gravity}
} 
\begin{align}
- \sum_{\alpha,\beta \in \mathcal{J}} \frac{\lambda_{\alpha \beta}}{6 |\mathcal{J}|^3}
 \sum_{\alpha_1,\ldots,\delta_3 \in \mathcal{J}}
& A_{\alpha_1 \beta_1 \gamma_1 \delta_1}
 A_{\alpha_2 \beta_2 \gamma_2 \delta_2}
 A_{\alpha_3 \beta_3 \gamma_3 \delta_3} \, \nn \\
& \times \, p_\alpha^{\delta_1 \alpha_2} p_\alpha^{\delta_2 \alpha_3}
 p_\alpha^{\delta_3 \alpha_1}
 p_\beta^{\beta_3 \gamma_2} p_\beta^{\beta_2 \gamma_1}
 p_\beta^{\beta_1 \gamma_3}, 
\label{3simplex_coloring_int} 
\end{align}
where $\lambda_{\alpha\beta}=\lambda_{\beta\alpha}$, 
and $p_\alpha$ is the projection matrix 
to the $\alpha$-th component:
\begin{align}
p_{\alpha}^{\alpha_1 \alpha_2} = \delta^{\alpha_1}_{\alpha} \, \delta^{\alpha_2}_{\alpha} .
\label{proj_matrix} 
\end{align}
The interaction terms can be expressed by thickened triangles 
as in Fig.~\ref{fig:3simplex_coloring_fig}, 
where the projection matrices $p_\alpha$ and $p_\beta$ are inserted to the index lines 
such that each side of the triangle has its own color. 
Thus,  each index triangle at a corner of a tetrahedron 
gives a factor of the form 
$\mathrm{tr}(p_{\alpha_1} p_{\alpha_2} p_{\alpha_3})$  
if there meet three triangles with colors 
$\alpha_1$, $\alpha_2$, $\alpha_3$ 
at the corner (see Fig.~\ref{fig:tetra_index_triangle_fig}). 
Since there are four corners in a tetrahedron, 
the tetrahedron illustrated in Fig.~\ref{fig:tetra_index_triangle_fig} gives the factor 
\begin{align}
\mathrm{tr}(p_{\alpha_1} p_{\alpha_2} p_{\alpha_3})\,
\mathrm{tr}(p_{\alpha_2} p_{\alpha_1} p_{\alpha_4})\,
\mathrm{tr}(p_{\alpha_1} p_{\alpha_3} p_{\alpha_4})\,
\mathrm{tr}(p_{\alpha_3} p_{\alpha_2} p_{\alpha_4}) \nn \\
 = \left\{ \begin{array}{l} 1 \quad (\alpha_1 = \alpha_2 = \alpha_3 = \alpha_4) \\ 0 \quad 
(\rm{otherwise}) \end{array} \right. .
\end{align}
This means that the index function $\mathcal{F}(\gamma)$ can take nonvanishing values 
only when four index triangles of each tetrahedron have the same color (say, $\alpha$), 
which enables us to say that the tetrahedron has a definite color $\alpha$. 
We thus succeed in coloring tetrahedra in $\gamma$. 
The parameters $\lambda_{\alpha \beta}$ in \eqref{3simplex_coloring_int} 
represent the coupling constants of local interactions 
among matter degrees of freedom on tetrahedra, 
because $\lambda_{\alpha\beta}$ appears in  $\mathcal{F}(\gamma)$ 
when the corresponding triangle is shared by neighboring tetrahedra 
of colors $\alpha$ and $\beta$. 

If we take the set of colors to be $\mathcal{J} = \mathbb{R}^D=\{{\bf x}\}$ 
and let the coupling constants 
$\lambda_{{\bf x},{\bf y}}$ $({\bf x},{\bf y} \in \mathbb{R}^D)$ 
take nonvanishing values only around ${\bf y}$ as a function of ${\bf x}$, 
then ${\bf x}$ can be interpreted as the target space coordinates 
of a tetrahedron in $\mathbb{R}^D$. 
Since neighboring tetrahedra are locally connected in $\mathbb{R}^D$, 
the model can describe the dynamics of membranes in $\mathbb{R}^D$. 
Instead, if we take $\mathcal{J}$  to be a finite set with $|\mathcal{J}|=q$, 
then the model can describe a $q$-state spin system on random volumes. 
In particular, if we consider the case $q = 2$ (with colors $\alpha = \pm$), 
then the model represents three-dimensional quantum gravity 
coupled to the Ising model. 
The system is ferromagnetic 
when $\lambda_{++} \geq \lambda_{+-}$ 
and $\lambda_{--} \geq \lambda_{+-}$. 
If the global $\mathbb{Z}_2$ symmetry ($+ \leftrightarrow -$) is explicitly broken 
by setting $\lambda_{++} \neq \lambda_{--}$, 
then the model describes a system in the presence of an external magnetic field.
With generic $q$, we can construct the $q$-state Potts models 
or the RSOS models \cite{Itzykson:1989sx} on random volumes 
by appropriately choosing $\lambda_{\alpha \beta}$.

\subsection{Coloring triangles}
\label{2simplex_coloring}

\begin{figure}[htbp]
\begin{center}
\includegraphics[height = 3.7cm]{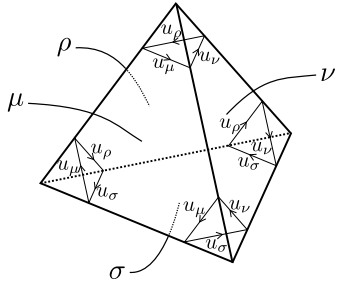}
\caption{
Index triangles inside a tetrahedron formed by colored triangles. 
}
\label{fig:2simplex_coloring_fig}
\vspace{-3ex}
\end{center}
\end{figure}

Triangles can be colored  
by making an argument similar to the one in subsection \ref{3simplex_coloring}. 
We set $\mathcal{A}_{\rm mat} = M_{s}(\mathbb{R})
=\bigoplus_{\alpha,\beta=1}^s
 \mathbb{R}\,e_{\alpha\beta}$,%
\footnote{
The linear dimension $s$ can be set to any value 
as long as the coupling constants \eqref{2simplex_coloring_factor} 
take desired forms. 
} 
and let the interaction terms take the form 
\begin{align}
- \sum_{\mu \in \mathcal{J}} \frac{\lambda_{\mu}}{6 s^2}
 A_{\alpha_1 \beta_1 \gamma_1 \delta_1}
 A_{\alpha_2 \beta_2 \gamma_2 \delta_2}
 A_{\alpha_3 \beta_3 \gamma_3 \delta_3}\,
 u_\mu^{\delta_1 \alpha_2} u_\mu^{\delta_2 \alpha_3}
 u_\mu^{\delta_3 \alpha_1} u_\mu^{\beta_3 \gamma_2}
 u_\mu^{\beta_2 \gamma_1} u_\mu^{\beta_1 \gamma_3} 
\label{2simplex_coloring_int}.
\end{align}
The model with \eqref{2simplex_coloring_int} generates diagrams 
where a color $\mu$ ($\mu \in \mathcal{J}$) is assigned to each triangle. 
If three triangles (with colors $\mu$, $\nu$, $\rho$) 
meet at a corner of a tetrahedron to construct an index triangle, 
the index function gets the factor $\mathrm{tr}(u_\mu u_\nu u_\rho)$.
Then, a tetrahedron illustrated in Fig.~\ref{fig:2simplex_coloring_fig} gives a factor of the form 
\begin{align}
 \mathrm{tr}(u_\mu u_\nu u_\rho)\,\mathrm{tr}(u_\nu u_\mu u_\sigma)\,
 \mathrm{tr}(u_\mu u_\rho u_\sigma)\,\mathrm{tr}(u_\rho u_\nu u_\sigma) . 
\label{2simplex_coloring_factor}
\end{align}
Such factors behave as the coupling constants 
of local interactions among matter degrees of freedom located on triangles.

There is another prescription to assign colors to triangles. 
We introduce $|\mathcal{J}|$ copies of variables $A$ and $B$ 
[denoted by $A^{(r)}$ and $B^{(r)}$ $(r=1,\ldots,|\mathcal{J}|)$], 
and let the action take the form 
\begin{align}
S =& \sum_{r \in \mathcal{J}} \frac{1}{2} A^{(r)}_{ij}B_{(r)}^{ij} 
- \sum_{r \in \mathcal{J}} \frac{\lambda_{r}}{6} 
A^{(r)}_{ij}A^{(r)}_{kl}A^{(r)}_{mn}\,g^{jk}g^{lm}g^{ni} 
\nonumber \\
&- \sum_{r_1 \ldots r_k \in \mathcal{J}} \sum_{k \geq 2} 
\frac{\mu_k^{r_1 \ldots r_k}}{2k} \,
B_{(r_1)}^{i_1 j_1} \cdots B_{(r_k)}^{i_k j_k}\, y_{i_1 \ldots i_k} y_{j_k \ldots j_1} 
\label{2simplex_coloring2_action}.
\end{align}
Then, this model also generates tetrahedral decompositions with colored triangles. 
The index function $\mathcal{F}(\gamma)$ 
of the model \eqref{2simplex_coloring2_action} 
gets the factor $\mu_k^{r_1 \ldots r_k}$ 
from a $k$-hinge shared by $k$ triangles with colors $r_1, \ldots, r_k$, 
and thus has a form different from \eqref{2simplex_coloring_factor}. 
Therefore, although matter degrees of freedom are assigned to triangles 
in both \eqref{2simplex_coloring_int} and \eqref{2simplex_coloring2_action}, 
they give different local interactions 
(at least apparently).

\subsection{Coloring edges}
\label{1simplex_coloring}

There are two prescriptions to assign colors to edges 
as is the case in coloring triangles. 

As in the first prescription in subsection \ref{2simplex_coloring}, 
we take $\mathcal{A}_{\rm mat} = M_{s}(\mathbb{R})
=\bigoplus_{\alpha,\beta=1}^s \mathbb{R}\,e_{\alpha\beta}$. 
We now let the interaction terms corresponding to hinges take the form 
\begin{align}
- \sum_{m \in \mathcal{J}} \sum_{k \geq 2} \frac{s^2 \mu_k^m}{2k}
 B^{\alpha_1 \beta_1 \gamma_1 \delta_1} \cdots
 B^{\alpha_k \beta_k \gamma_k \delta_k} \,
 u^{m}_{\beta_1 \alpha_2} \ldots u^{m}_{\beta_k \alpha_1}
 u^{m}_{\gamma_1 \delta_2} \ldots u^{m}_{\gamma_k \delta_1}. 
\label{1simplex_coloring1_int}
\end{align}
This generates diagrams where each edge has a color $m$ ($m \in \mathcal{J}$), 
and each index triangle gives the factor 
$\mathrm{tr}(u^{m_1} u^{m_2} u^{m_3})$ depending on the colors of the edges.
They give the coupling constants of local interactions 
among matter degrees of freedom located on edges. 

Another prescription to assign colors to edges 
can be given by modifying the action \eqref{original_action} to the form 
\begin{align}
 S =& \sum_{r \in \mathcal{J}} \frac{1}{2} A^{(r)}_{ij}B_{(r)}^{ij} 
 - \sum_{r_1,r_2,r_3 \in \mathcal{J}}
 \frac{\lambda_{r_1 r_2 r_3}}{6}  A^{(r_1)}_{ij}A^{(r_2)}_{kl}A^{(r_3)}_{mn}g^{jk}g^{lm}g^{ni} \nonumber 
\\
&- \sum_{r \in \mathcal{J}} \sum_{k \geq 2} \frac{\mu_k}{2k} 
 B_{(r)}^{i_1 j_1} \ldots B_{(r)}^{i_k j_k} y_{i_1 \ldots i_k} y_{j_k \ldots j_1} 
\label{1simplex_coloring2_action}.
\end{align}
Each triangle gives the factor $\lambda_{r_1 r_2 r_3}$ 
if three hinges (with colors $r_1$, $r_2$, $r_3$) meet there.

\subsection{Coloring vertices}
\label{0simplex_coloring}

Vertices can also be colored 
despite the fact that the action \eqref{original_action} does not have interaction terms 
corresponding to vertices. 

We first set the matter associative algebra 
to be $\mathcal{A}_{\rm mat}
 = \mathcal{A}_{\rm mat}^{(1)} \oplus \ldots
 \oplus \mathcal{A}_{\rm mat}^{(|\mathcal{J}|)}$, 
and let the interaction terms corresponding to hinges take the form
\begin{align}
- \sum_{\alpha, \beta \in \mathcal{J}} \sum_{k \geq 2} \frac{\mu_k^{\alpha \beta}}{2k}\,
 B^{i_1 j_1} \ldots B^{i_k j_k} y^{(\alpha)}_{i_1 \ldots i_k}\,
 y^{(\beta)}_{j_k \ldots j_1} .
\label{0simplex_coloring1_int}
\end{align}
Here $y^{(\alpha)}_{i_1 \ldots i_k}$ are the coupling constants 
constructed from the structure constants $y_{\,\,\,ij}^{(\alpha)\, k}$ 
of $A_{\rm mat}^{(\alpha)}$ 
and take nonvanishing values 
only when all the indices $i_1, \ldots, i_k$ belong to 
$\mathcal{A}_{\rm mat}^{(\alpha)}$. 
Accordingly, all the junctions in the same connected index network 
should have the same color $\alpha$ 
in order for the index function $\mathcal{F}(\gamma)$ 
to take nonvanishing values. 
Thus, we can assign a color to the index network of each vertex in diagram $\gamma$, 
and can say that the model generates diagrams with colored vertices. 
The matter degrees of freedom located on vertices 
have local interactions, 
and two neighboring vertices with colors $\alpha$ and $\beta$ 
(connected by a hinge) 
gives the factor $\mu_k^{\alpha \beta}$ to $\mathcal{F}(\gamma)$. 

The above coloring of vertices can also be realized 
by setting $\mathcal{A}_{\rm mat} = M_{|\mathcal{J}|}(\mathbb{R})$ 
and letting the interaction terms corresponding to hinges take the form
\begin{align}
 - \sum_{\alpha, \beta \in \mathcal{J}} \sum_{k \geq 2} 
 \frac{|\mathcal{J}|^2 \mu_k^{\alpha \beta}}{2k}\,
 B^{\alpha_1 \beta_1 \gamma_1 \delta_1} \cdots
 B^{\alpha_k \beta_k \gamma_k \delta_k}
 p^{\alpha}_{\beta_1 \alpha_2} \cdots p^{\alpha}_{\beta_k \alpha_1}
 p^{\beta}_{\gamma_1 \delta_2} \cdots p^{\beta}_{\gamma_k \delta_1} 
\label{0simplex_coloring2_int}, 
\end{align}
where $p^\alpha$ is the projection matrix to the $\alpha$-th component 
[the same as the one given in \eqref{proj_matrix}]. 
It is easy to see that 
this model is dual to the model with \eqref{3simplex_coloring_int}
through the duality transformation \eqref{duality_trsf}.  
That is, the action with the interaction term \eqref{0simplex_coloring2_int} 
can be regarded as a $q$-state system on the dual lattice of $\gamma$ 
($q=|\mathcal{J}|$). 

We thus conclude that 
matter degrees of freedom can be introduced to triangle-hinge models 
such that they live on simplices of any dimensions 
and interact with themselves locally.

\section{Relations to colored tensor models}
\label{ctm}

We can further construct various kinds of models 
by combining several prescriptions explained in the previous section. 
For example, we show in this section that 
three-dimensional colored tensor models \cite{Gurau:2011xp} 
can be realized as triangle-hinge models 
by coloring tetrahedra, triangles and edges {\em at a time\,}.

\subsection{Feynman rules of colored tensor models}
\label{ctm_review}

\begin{figure}[htbp]
\begin{center}
\includegraphics[height = 3.0cm]{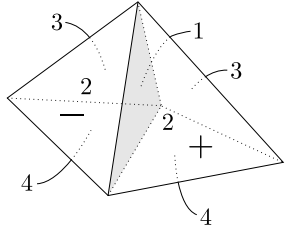}
\caption{
A part of a Feynman diagram in colored tensor models. 
There are two tetrahedra, 
one corresponding to an interaction vertex proportional to $\kappa$ 
and the other to $\bar{\kappa}$. 
The two adjacent tetrahedra have opposite orientations. 
}
\label{fig:ctm_diagram_fig}
\vspace{-3ex}
\end{center}
\end{figure}

We first review the Feynman rules of three-dimensional colored tensor models 
(see, e.g., \cite{Gurau:2011xp} for a review). 
The dynamical variables of colored tensor models are given 
by a pair of rank-three tensors $\phi^\mu_{IJK}$ and $\bar{\phi}^\mu_{IJK}$ 
with no symmetry properties under permutations of the subscripts $I, J, K$. 
The tensors represent two kinds of colored triangles, 
where $\{I\}$ is the set of indices assigned to edges, 
and $\{\mu\}=\{1,2,3,4\}$ are the colors assigned to triangles.%
\footnote{
In the original Boulatov model \cite{Boulatov:1992vp}
the index $I$ runs over the elements of group manifold $SU(2)$. 
} 
The action takes the form 
\begin{align}
 S = \sum_{\mu = 1}^4 \phi^\mu_{IJK}\bar{\phi}^\mu_{IJK}
 + \kappa\, \phi^1_{IJK} \phi^2_{KML} \phi^3_{MJN} \phi^4_{LNI}
 + \bar{\kappa}\, \bar{\phi}^1_{IJK} \bar{\phi}^2_{KML}
 \bar{\phi}^3_{MJN} \bar{\phi}^4_{LNI} 
\label{ctm_action}.
\end{align}
Looking at the way of contraction of indices $I$, 
one easily sees that this action generates the Feynman diagrams 
where the interaction vertices can be identified with tetrahedra 
which are glued at their faces through the propagator. 
Since there are two types of interaction terms 
$\kappa\,\phi^4$ and $\bar{\kappa}\,\bar{\phi}^4$, 
the set of tetrahedra can be decomposed to two different classes, 
which we label with $\alpha=\pm$, respectively. 
We assign four different colors $\mu=1,\ldots,4$  
to four triangles of each tetrahedron. 
This coloring of triangles naturally introduces 
the coloring of six edges in a tetrahedron, 
and we assign color $(\mu\nu)=(\nu\mu)$ to an edge 
if the edge is shared by two triangles with colors $\mu$ and $\nu$ $(\mu\neq \nu)$. 
Since the tensors $\phi^\mu_{IJK}$ and $\bar{\phi}^\mu_{IJK}$ 
have no permutation symmetry with respect to the subscripts, 
two tetrahedra can be glued at their faces 
only when two triangles to be identified have the same color $\mu$ 
and two edges to be identified have the same color $(\mu\nu)$ 
as in Fig.~\ref{fig:ctm_diagram_fig}. 
We say that the tetrahedron has positive (or negative) orientation 
if triangles $1,2,3$ are located clockwise (or counterclockwise) 
when seen from triangle 4 (see Fig.~\ref{fig:ctm_diagram_fig}). 
Since the kinetic term has the form $\phi\,\bar{\phi}$ 
(not including $\phi^2$ or $\bar{\phi}^2$), 
two adjacent tetrahedra must have different orientations.

The Feynman rules for colored tensor models \eqref{ctm_action}
thus can be summarized as follows: 
\begin{enumerate}
\item
Interaction vertices are represented by two types (orientations) of tetrahedra, 
$\alpha=\pm$, 
and any two adjacent tetrahedra have different types. 
\item
Four different colors $\mu=1,\ldots,4$ are assigned  
to four triangles of each tetrahedron,  
such that the assignment agrees with 
the orientation of the tetrahedron when $\alpha=+$, 
while it is opposite when $\alpha=-$. 
\item
Two tetrahedra are glued at their faces 
in such a way that two triangles to be identified have the same color $\mu$ 
and two edges to be identified have the same color $(\mu\nu)$. 
\end{enumerate}

\subsection{Realization of colored tensor models 
as triangle-hinge models}
\label{ctm_matter}

\begin{figure}[htbp]
\begin{center}
\includegraphics[height = 3.0cm]{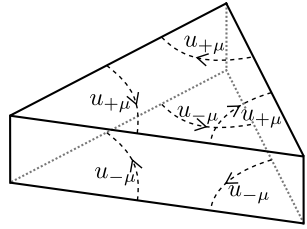}
\caption{
A thickened triangle vertex 
coming from the action \eqref{t-h2ctm_action}, 
which realizes colored tensor models.
}
\label{fig:ctm_coloring_int_fig}
\vspace{-3ex}
\end{center}
\end{figure}

The above Feynman rules for three-dimensional colored tensor models 
can be reproduced from triangle-hinge models 
by coloring tetrahedra, triangles and edges at a time. 
To see this, we set the matter associative algebra $\mathcal{A}_{\rm mat}$
to be a matrix ring $M_{2s}(\mathbb{R})$ 
and let the action take the form 
\begin{align}
S = & \sum_{(\mu \nu)} \frac{1}{2} A_{\alpha \beta \gamma \delta}^{(\mu \nu)}
B^{\alpha \beta \gamma \delta}_{(\mu \nu)}
 - \frac{\lambda}{6 (2s)^3} \sum_{\mu = 1}^{4}
 \,\frac{1}{6}\!\!
 \sum^4_{\substack{\nu,\rho,\sigma=1 \\(\mu\nu\rho\sigma) :{\rm \,\,all\,\,different}}}
 A_{\alpha_1 \beta_1 \gamma_1 \delta_1}^{(\mu \nu)} 
 A_{\alpha_2 \beta_2 \gamma_2 \delta_2}^{(\mu \rho)} 
 A_{\alpha_3 \beta_3 \gamma_3 \delta_3}^{(\mu \sigma)} 
\nonumber\\
 &~~~\times\,\bigl(u_{+\mu}^{\delta_1 \alpha_2} u_{+\mu}^{\delta_2 \alpha_3} 
 u_{+\mu}^{\delta_3 \alpha_1} u_{-\mu}^{\beta_3 \gamma_2}
 u_{-\mu}^{\beta_2 \gamma_1} u_{-\mu}^{\beta_1 \gamma_3}
 + u_{-\mu}^{\delta_1 \alpha_2} u_{-\mu}^{\delta_2 \alpha_3}
 u_{-\mu}^{\delta_3 \alpha_1} u_{+\mu}^{\beta_3 \gamma_2}
 u_{+\mu}^{\beta_2 \gamma_1} u_{+\mu}^{\beta_1 \gamma_3} \bigr)
\nonumber\\
 &~~~ - \sum_{k \geq 2} \frac{n^2 \mu_k}{2k} \sum_{(\mu \nu)}
 B^{\alpha_1 \alpha_2 \beta_2 \beta_1}_{(\mu \nu)} B^{\alpha_2 \alpha_3 \beta_3 \beta_2}_{(\mu \nu)}
 \cdots B^{\alpha_k \alpha_1 \beta_1 \beta_k}_{(\mu \nu)}. 
\label{t-h2ctm_action}
\end{align}
Here, the indices $(\mu \nu) = (\nu \mu)\ (\mu,\nu=1,\ldots,4; \ \mu \neq \nu)$ 
stand for the colors assigned to edges, 
the sum $\sum_{(\mu \nu)}$ is taken over all different colors of edges, 
and we again have neglected the gravity part 
which ensures the resulting Feynman diagrams 
to form a set of tetrahedra (see footnote \ref{fn_gravity}). 
We further assume the matrices $u_{\pm \mu}$ to have the form\begin{align}
 &u_{+\mu} = \biggl( \begin{array}{cccc}u_\mu&&0\\0&&0 \end{array}\biggr),
 \quad
 u_{-\mu} = \biggl( \begin{array}{cccc}0&&0\\0&&
 (u_\mu)^{\mathrm{T}} \end{array}\biggr). 
\end{align}
\newpage
\noindent
Here $s\times s$ matrices $u_\mu$ are chosen 
such that they satisfy%
\footnote{
For example, one can take the following $6\times 6$ matrices: 
\begin{align}
 u_1 = 2^{-\frac{2}{3}} \left( \begin{array}{cccccc} 0&&\sigma_1&&0 
\\
 0&&0&&\sigma_1 \\ 0&&0&&0 \end{array} \right), \quad
 u_2 = 2^{-\frac{2}{3}} \left( \begin{array}{cccccc} 0&&0&&0 
\\
 0&&0&&-i\sigma_2 \\ -i\sigma_2&&0&&0 \end{array} \right), 
\nonumber \\
 u_3 = 2^{-\frac{2}{3}} \left( \begin{array}{cccccc} 0&&\sigma_3&&0 \\ 0&&0&&0 
\\
 \sigma_3&&0&&0 \end{array} \right), \quad
 u_4 = 2^{\frac{1}{3}} \left( \begin{array}{cccccc} 1&&i\sigma_2&&0 \\ 0&&1&&-\sigma_3 
\\
 -\sigma_1&&0&&1 \end{array} \right), \nonumber
\end{align}
where $\sigma_i$ $(i = 1,2,3)$ are the Pauli matrices. 
}
\begin{align}
 &\mathrm{tr}(u_\mu u_\nu u_\rho)\,\mathrm{tr}(u_\nu u_\mu u_\sigma)
 \,\mathrm{tr}(u_\mu u_\rho u_\sigma)\,\mathrm{tr}(u_\rho u_\nu u_\sigma)
 = \biggl\{ 
 \begin{array}{ll} 1 & (\epsilon_{\mu \nu \rho \sigma} = +1) \\
 0 & (\rm{otherwise}) \end{array}, 
\label{tetra_index_triangle}
\end{align}
where $\epsilon_{\mu \nu \rho \sigma}$ is the totally antisymmetric tensor 
with $\epsilon_{1234}=1$. 
The interaction vertices corresponding to triangles 
can be expressed by thickened triangles as in Fig.~\ref{fig:ctm_coloring_int_fig}.
Note that we make colorings for simplices of three different dimensions 
(tetrahedra, triangles and edges), 
which are described in subsections \ref{3simplex_coloring}, 
\ref{2simplex_coloring} and \ref{1simplex_coloring}, respectively. 
In fact, each tetrahedron has a type (orientation) $\alpha=\pm$, 
each triangle has a color $\mu=1,\ldots,4$, 
and each edge has a color $(\mu\nu)=(\nu\mu)$ $(\mu\neq\nu)$. 
The interaction terms corresponding to triangles indicate that 
the three edges of a triangle of color $\mu$ have 
different colors $(\mu \nu)$, $(\mu \rho)$, $(\mu \sigma)$. 
Note that we particularly set $\lambda_{\alpha\beta}$ as 
$\lambda_{++}=\lambda_{--}=0$ 
(and $\lambda_{+-}=\lambda_{-+}= \lambda$), 
so that any two adjacent tetrahedra have different types. 
As can be seen from \eqref{2simplex_coloring_factor},
a tetrahedron of type $\alpha=+$ (or $\alpha=-$) gives the factor
\begin{align}
 \mathrm{tr}\bigl(u_{\alpha\mu} u_{\alpha\nu} u_{\alpha\rho}\bigr)
 \mathrm{tr}\bigl(u_{\alpha\nu} u_{\alpha\mu} u_{\alpha\sigma}\bigr)
 \mathrm{tr}\bigl(u_{\alpha\mu} u_{\alpha\rho} u_{\alpha\sigma}\bigr)
 \mathrm{tr}\bigl(u_{\alpha\rho} u_{\alpha\nu} u_{\alpha\sigma}\bigr) 
 \quad (\alpha=\pm),
\end{align}
which takes a nonvanishing value $(=1)$ 
only when the four colors $\mu,\nu,\rho,\sigma$ are all different 
and correspond to 
the positive (or negative) orientation. 
Thus, a tetrahedron in a nonvanishing Feynman diagram 
has a positive orientation when its color is $\alpha=+$ 
and has a negative orientation when $\alpha=-$. 
Furthermore, if two triangles sharing an edge of color $(\mu\nu)$ $(\mu\neq\nu)$
belong to the same tetrahedron, 
then one of the two triangles has color $\mu$ 
and the other has color $\nu$. 
In fact, any triangle connected to a hinge of color $(\mu\nu)$ 
must have color $\mu$ or $\nu$, 
but two triangles sharing the edge $(\mu\nu)$ 
must have different colors 
if they belong to the same tetrahedron. 

Therefore, the Feynman diagrams generated by the action \eqref{t-h2ctm_action} 
consist of tetrahedra 
where two adjacent tetrahedra have different orientations $\alpha=\pm$, 
and four triangles in each tetrahedron have different colors $\mu=1,\ldots,4$ 
such as to be consistent with the orientation of the tetrahedron. 
Furthermore, 
the coloring of each edge does not depend on the choice 
of a tetrahedron including the edge, 
which leads us to the interpretation 
that two tetrahedra are glued at their faces 
such that the edges to be identified have the same color. 
We thus conclude that the Feynman diagrams obtained from the action \eqref{t-h2ctm_action} 
obey the same Feynman rules 
obtained from the action \eqref{ctm_action} 
of three-dimensional colored tensor models.

\section{Conclusion}
\label{conclusion}
In this paper, we give a general prescription to introduce 
matter degrees of freedom to triangle-hinge models. 
This is achieved by setting the defining associative algebra $\mathcal{A}$ 
to be a tensor product of the form $\mathcal{A}_{\rm grav} \otimes \mathcal{A}_{\rm mat}$ 
and by modifying the interaction terms appropriately. 
The matter fields thus obtained have local interactions, 
since a colored tetrahedron can only interact with the neighbors. 
We can assign colors not only to tetrahedra  
but also to simplices of arbitrary dimensions. 
We further show that there exists a duality 
between matter fields on a tetrahedral lattice 
and those on its dual lattice, 
which can be realized by applying the duality transformation \eqref{duality_trsf} 
which interchanges the roles of triangles and hinges. 

When we take the set of colors to be $\mathbb{R}^D$ 
and assign colors to tetrahedra as in \eqref{3simplex_coloring_int}, 
the matter fields represent the target space coordinates 
of membranes in $D$ dimensions. 
By taking different sets of colors, 
we can also construct various spin systems on random volumes, 
including the Ising model, the $q$-state Potts models and the RSOS models. 

A wider class of models can be further obtained as triangle-hinge models 
by coloring lower dimensional simplices as well as tetrahedra. 
For example, three-dimensional colored tensor models 
can be obtained by coloring tetrahedra, triangles and edges at a time. 
This is shown in section \ref{ctm_matter} 
by explicitly demonstrating that the same Feynman rules are obtained. 

It should be interesting to investigate the critical behaviors 
of triangle-hinge models with matter fields. 
In particular, it is important to study the case 
when matter fields correspond to the target space coordinates 
of embedded membranes. 
It is also interesting to investigate 
if there is any obstacle in introducing matter fields 
like the ``$c=1$ barrier'' for matter fields on random surfaces. 
Introduction of supersymmetry to triangle-hinge models 
is another interesting problem. 
Studies in these directions are now in progress 
and will be communicated elsewhere. 

\begin{acknowledgments}
MF is supported by MEXT (Grant No.\,23540304).
SS is supported by the JSPS fellowship.
\end{acknowledgments}

\bibliography{draft_v3}

\end{document}